\documentclass[twocolumn,amsmath,amssymb]{revtex4}

\usepackage{epsfig,wrapfig}
\usepackage{dcolumn}
\usepackage{bm}

\newcommand{\beq}{\begin{equation}}
\newcommand{\eeq}{\end{equation}}

\begin{document}

\title{
Spin degrees of freedom and flattening of the spectra
of single-particle \\ excitations in strongly correlated
Fermi systems
}

\author{
V.~A.~Khodel$^{1}$, P.~Schuck$^{2}$, and M.~V.~Zverev$^{1}$  }

\affiliation {%
${}^{1}$  Kurchatov Institute, Russian Research Center,
    Moscow, 123182 Russia \\
${}^{2}$ Groupe de Physique Th\'{e}orique, Institut de Physique
       Nucl\'{e}aire,\\
     F-01406 Orsay Cedex, France
}

\date{\today}

\begin{abstract}
The impact of long-range spin-spin correlations on the structure
of a flat portion in single-particle
spectra $\xi(p)$, which emerges beyond the point, where
 the Landau state loses its stability,  is studied. We  supplement
the well-known Nozieres model of a Fermi system with
limited scalar long-range forces by a similar long-range
spin-dependent term and calculate the spectra versus its strength
$g$. It is found that Nozieres results hold as long as $g>0$.
However, with $g$ changing its sign,
the spontaneous magnetization is shown to arise at any nonzero $g$.
The increase of the strength $|g|$ is demonstrated to result in
shrinkage of the domain in momentum space, occupied by
the flat portion of $\xi(p)$, and, eventually, in its vanishing.
\end{abstract}

\maketitle

The investigation of flattening of single-particle (sp) spectra
of Fermi liquids is dated back to
Ref.~\cite{don}, where  long-range correlations,
enhanced  in the vicinity of an impending ferromagnetic phase
transition, were shown to result in the divergence of
 the effective mass  $M^*$ at the  transition point.
Later in \cite{ks} an idea of the so-called fermion condensation,
i.e. a rearrangement of the Landau state, occurring beyond a critical
point in strongly correlated Fermi systems with long-range effective
interactions, was suggested.
A striking feature of this rearrangement is
"swelling" of the Fermi surface (FS), i.e. the occurrence of
a completely flat portion $\xi({\bf p})=0$,
called the fermion condensate (FC), in the spectrum $\xi({\bf p})$,
measured from the FS.

To gain insight into the problem of fermion condensation, let us turn
to the Dyson  equation, rewriting it in the form
\beq
\xi({\bf p})=\xi^0_{{\bf p}}+\Sigma({\bf p},\xi({\bf p}))  \   ,
\label{sp1}
\eeq
 appropriate for finding the FC solutions $\xi({\bf p})=0$.
The sp mass operator $\Sigma$ is usually determined by the formula
\beq
\Sigma({\bf p},\varepsilon)=
\int W{\bf p},\varepsilon,{\bf p}_1,\varepsilon_1)
G({\bf p}_1,\varepsilon_1){d^4p_1\over (2\pi)^4 i}  \  ,
\label{gw}
\eeq
 where $W$ is an effective interaction between particles,
and  $G$ is the sp Green function. It is worth noting that
 the imaginary part of
$\Sigma(p,\varepsilon)$  vanishes at $\varepsilon=0$.
 Therefore, in searching for the FC solutions, only the real part
of $\Sigma$ is relevant.

 For a long time,  it was  reckoned
that  in homogeneous Fermi systems, there exists a one-to-one
correspondence between
the momentum $p$ and the sp energy $\xi$, at least, close to the FS,
i.e. the derivative $(d\xi/dp)_F$ is always positive, a postulate,
virtually being  a cornerstone of the Landau theory of Fermi liquid
\cite{lan}.
However, in systems with long-range forces it might  be incorrect.
Compelling evidence for that is provided
by a phenomenological  model, suggested by
Nozieres \cite{noz}, where $W$
is taken as a constant in the coordinate space. Thus, in this model
Eq.~(\ref{sp1}) takes the form
\beq
\xi-\xi^0_{{\bf p}}=f n(\xi)  \    ,
\label{eqnoz}
\eeq
where $f$ is an effective coupling constant, the value of which
is taken to be positive,  while
$n(\xi)$ is the Landau quasiparticle momentum distribution, being 0
at positive $\xi$, and 1, at negative ones.
This equation is easily solved, but we concentrate on its
graphical solution (see Fig.~1).
Let us draw both the r.h.s.
and the l.h.s. of Eq.~(\ref{eqnoz}) as functions of $\xi$,
taking $p$ as an input parameter.
The l.h.s.~of (\ref{eqnoz}), depicted at different $p$,
provides a set of parallel straight lines, while the r.h.s.
forms a kink,
the vertical segment of which is located at $\xi=0$,
no matter what is the input. Crossing points yield
the sp spectrum $\xi(p)$. If they lie on any of the two
horizontal pieces of the kink, $\xi(p)$ does coincide
with the sp spectrum of ideal Fermi gas. An
unconventional situation occurs when straight lines cross
the steep section of the kink.
In this case, the intersection point remains 0 in a finite momentum
interval  $[p_i,p_f]$.
\begin{figure}
\mbox{\epsfig{file=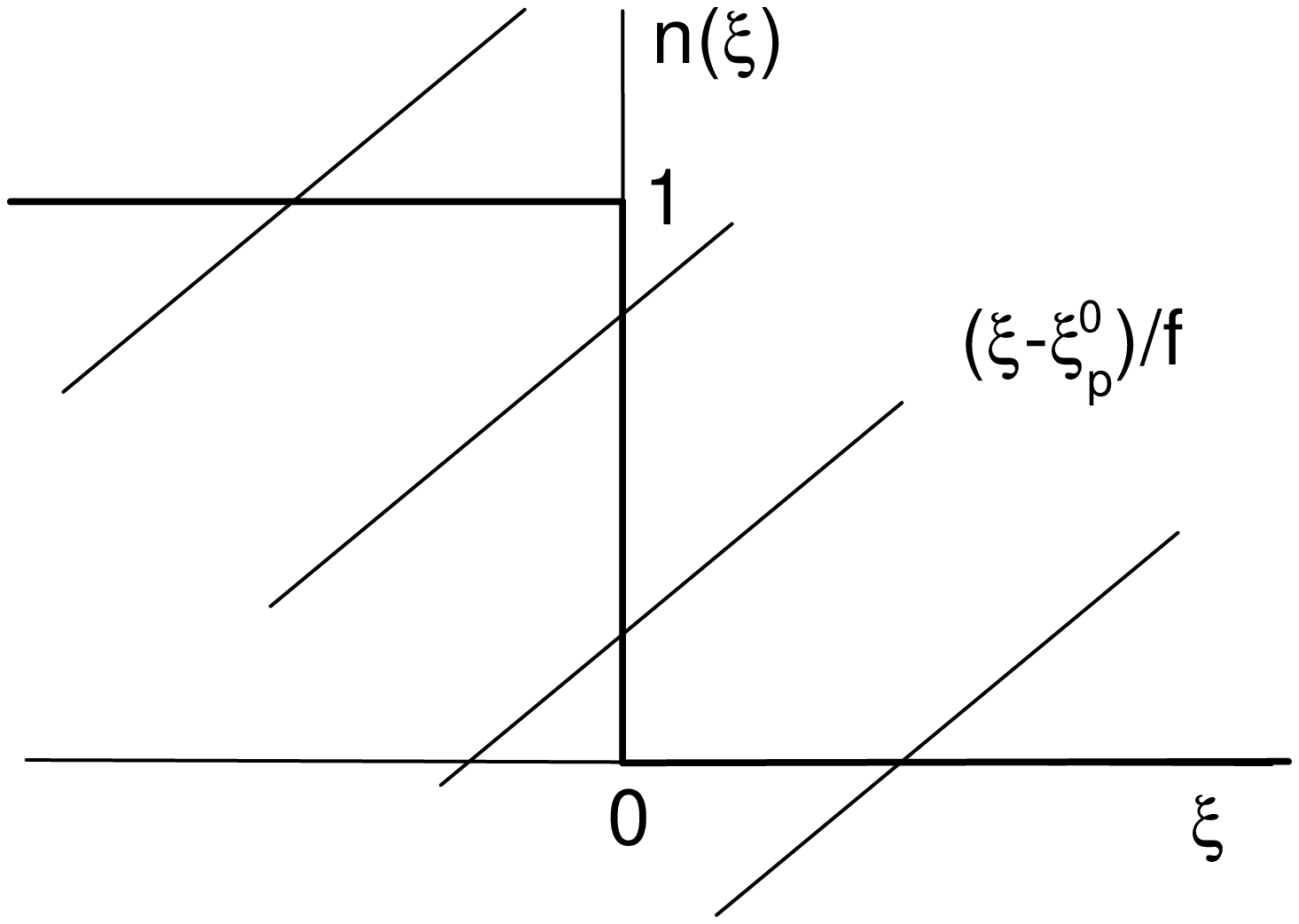,height=4.cm,width=6.cm}}
\caption{
Graphical illustration of the solution of
Eq.~(\ref{eqnoz}).
}
\end{figure}
This plateau $\xi(p)=0$,
lying exactly at the FS, does form the fermion condensate.
Since $\xi(p)=0$ in the finite volume of momentum space, the
density of states $\rho(\varepsilon)$  acquires a singular term
$\rho(\varepsilon)\sim\delta(\varepsilon)$. Another
salient feature of the fermion condensation phenomenon is that
in the FC domain,
the quasiparticle
momentum distribution $n_0(p)$ differs from the Landau one.
Indeed, upon  setting  $\xi=0$ in Eq.(\ref{eqnoz})
one finds
\beq
n_0(p)=-{\xi^0_{{\bf p}}\over f} \  ,\qquad   p_i<p<p_f \  ,
\label{qd}
\eeq
in contrast to the conventional step function $n_F(p)=\theta(p_F-p)$.
Thus, the FS does swell, whereas the basic assertion
of Landau theory fails.

In dealing with the FC problem, attention is usually paid to
the spin-independent part  of the effective interaction $W$
(see e.g. \cite{ks,noz,vol,kcs,schuck,zb}). Here we investigate
effects, associated with long-range spin-spin components of $W$,
which involve the spin-up quasiparticle
distribution $n_+(p)$ and spin-down one $n_-(p)$.
It is instructive to start such an analysis with a generalization
of the Nozieres model \cite{noz}, supplementing it by long-range
spin-spin terms. As a result, one obtains two equations
\begin{eqnarray}
\xi_+(p)&=&\xi^0_{{\bf p}} +{1\over 2}\,f\,(n_+(p)+n_-(p))
\nonumber \\ &&
+{1\over 2}\,g\,(n_+(p)-n_-(p)) \ ,
\nonumber \\
\xi_-(p)&=&\xi^0_{{\bf p}} +{1\over 2}\,f\,(n_+(p)+n_-(p))
\nonumber \\ &&
-{1\over 2}\,g\,(n_+(p)-n_-(p)) \   ,
\label{spin}
\end{eqnarray}
with a new constant  $g$, specifying  the long-range spin-spin
 component of the model  effective interaction $W$.
It is worth noting that in the case of spin fluctuations
with the nonzero critical momentum $q_c \ll p_F$,
presumably relevant to two-dimensional
liquid He-3, the constants in Eq.~(\ref{spin}) are related
with each other: $g=-f/3$.

In conventional Fermi liquids, the spontaneous spin $S$ arises
only if strength of the spin-spin interaction obeys the
Pomeranchuk condition $|G|\rho(0)>1$, where $\rho(0)=p_FM^*/\pi^2$
is the density of states at the FS in the Landau theory. In the
case at issue, the density of states is infinite, so  that one can
expect the emergence of the spontaneous magnetization at any
$g<0$.

\begin{figure}
\mbox{\epsfig{file=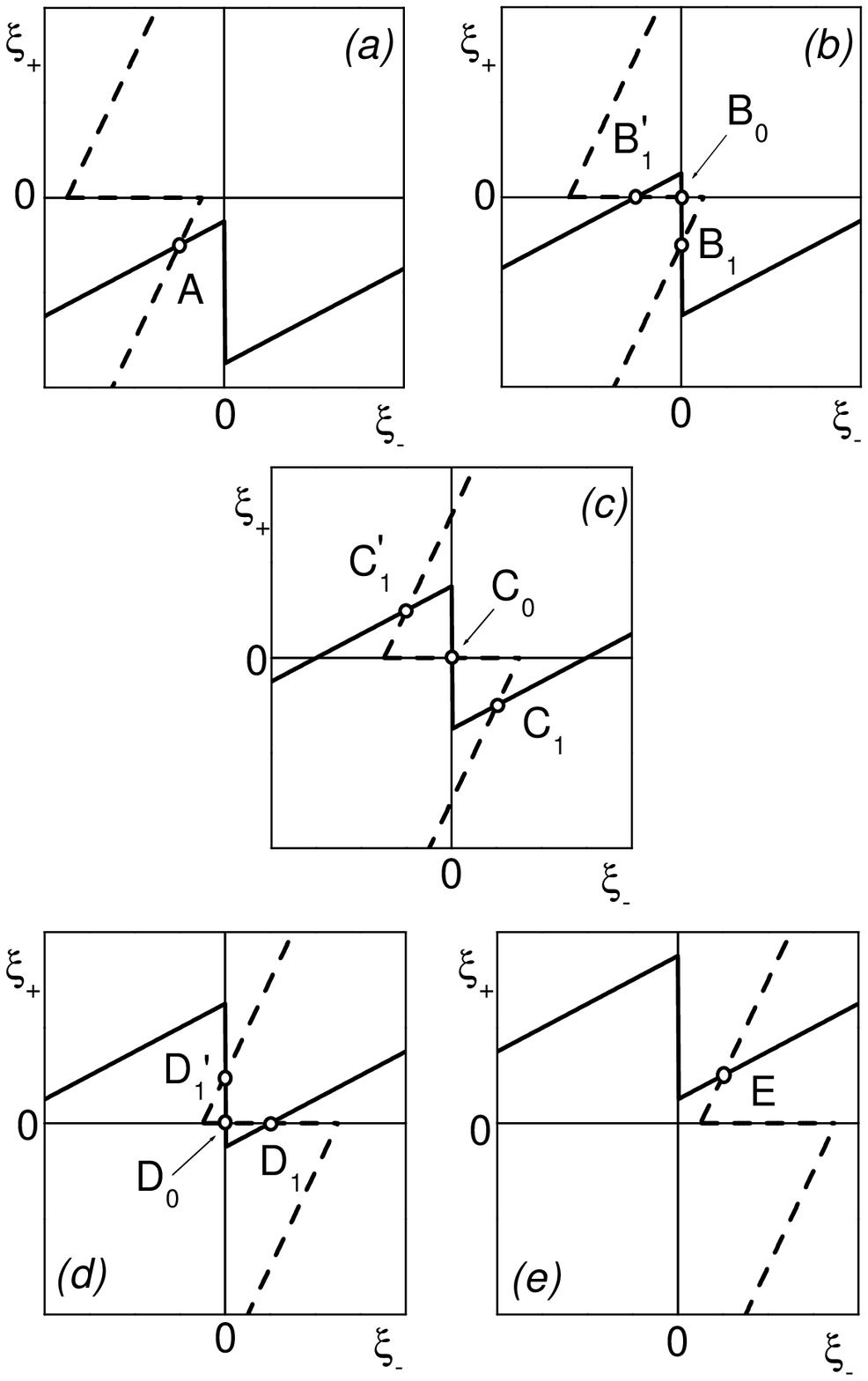,height=9.cm,width=6.cm}}
\caption{
Graphical illustration of the dynamics of solutions of
the set (\ref{syst2}) with increasing $\xi^0_{\bf p}$
in the case of $|g|<f$.
}
\end{figure}

To facilitate the solution of the problem, we recast
the system (\ref{spin}) to the form
\begin{eqnarray}
\xi_+&=&\left(1-{a\over b}\right)\xi^0_{{\bf p}}
+{a\over b}\xi_-+\left(b-{a^2\over b}\right)n_-\  ,
\nonumber \\
\xi_-&=&\left(1-{a\over b}\right)\xi^0_{{\bf p}} +{a\over b}\xi_ +
+\left(b-{a^2\over b}\right)n_+ \  ,
\label{syst2}
\end{eqnarray}
 where $a=(f+g)/2$ and $b=(f-g)/2$.
Let us now draw the plot $\xi_+(\xi_-)$, proceeding from the first of
Eqs.(\ref{syst2}) and treating $p$ as an input parameter.
At $\xi_->0$ the function $\xi_+(\xi_-)$  is given by  the straight
line $\xi_+= (1-a/ b)\xi^0_{{\bf p}} +a\xi_-/b$,
while at $\xi_-<0$ this straight line is slightly shifted:
$\xi_+= (1-a/ b)\xi^0_{{\bf p}}+(b^2-a^2)/b +a\xi_-/b$.
The drawing of its counterpart $\xi_-$ from the second of
Eqs.(\ref{syst2}) yields another couple of  straight lines:
$\xi_-= (1-a/ b)\xi^0_{{\bf p}} +a\xi_+/b$ at $\xi_+>0$,
and $\xi_-= (1-a/ b)\xi^0_{{\bf p}}-(b^2-a^2)/b +a\xi_-/b$
at $\xi_+<0$.  As we shall see, both these
curves cross each other either at one or at three points, providing
several solutions for the spectra $\xi_{\pm}(p)$.

\begin{figure}
\mbox{\epsfig{file=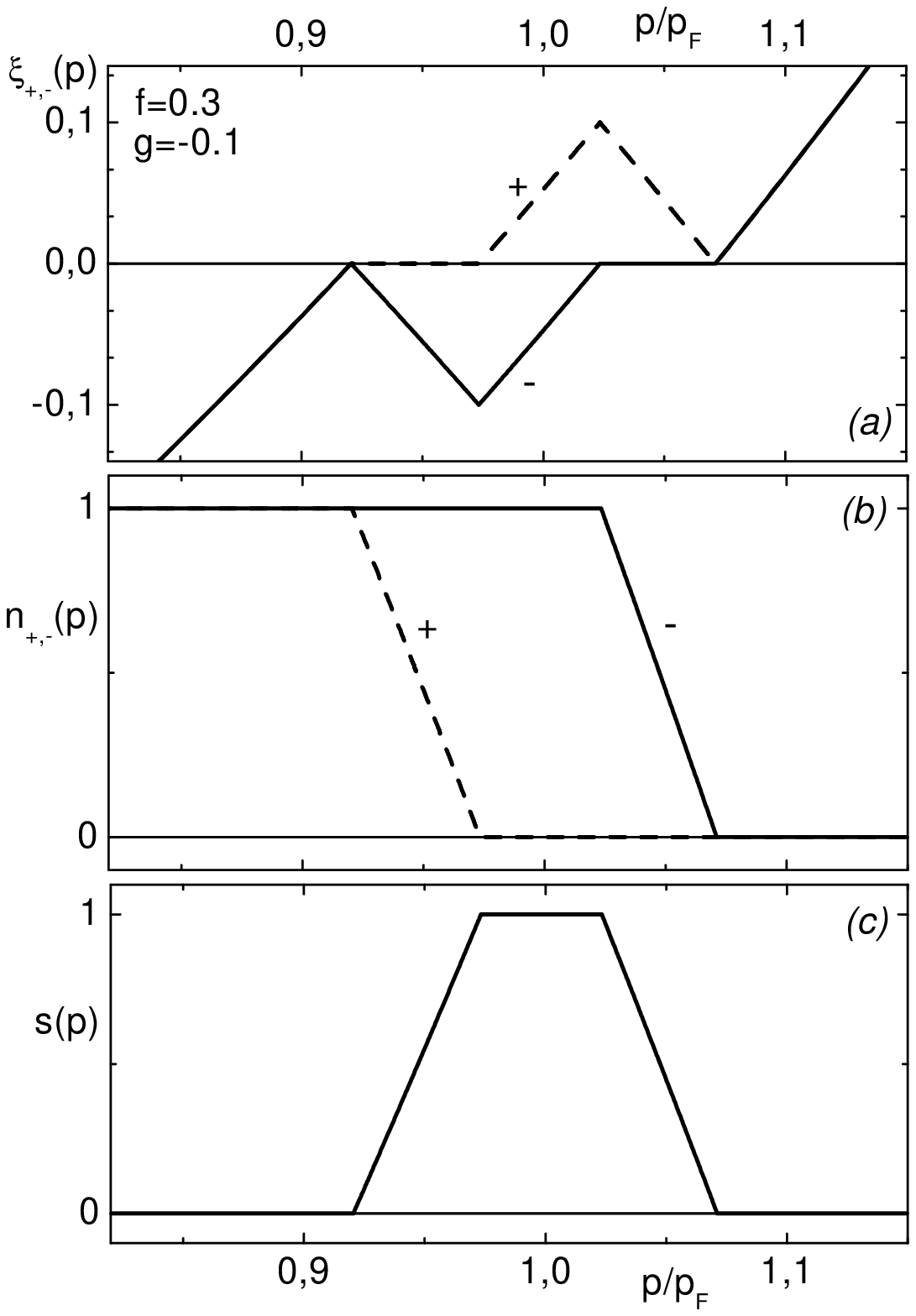,height=8.cm,width=4.5cm}}
\caption{
Single-particle spectra $\xi{\pm}(p)$ in units of
$\varepsilon^0_F=p_F^2/2M$ (panel (a)), occupation numbers
$n_{\pm}(p)$ (panel (b)) and spontaneous spin
$s(p)=n_-(p)-n_+(p)$ (panel (c)), calculated for
$f=0.3$, $g=-0.1$ in units of $\varepsilon^0_F$.
}
\end{figure}

\begin{figure}
\mbox{\epsfig{file=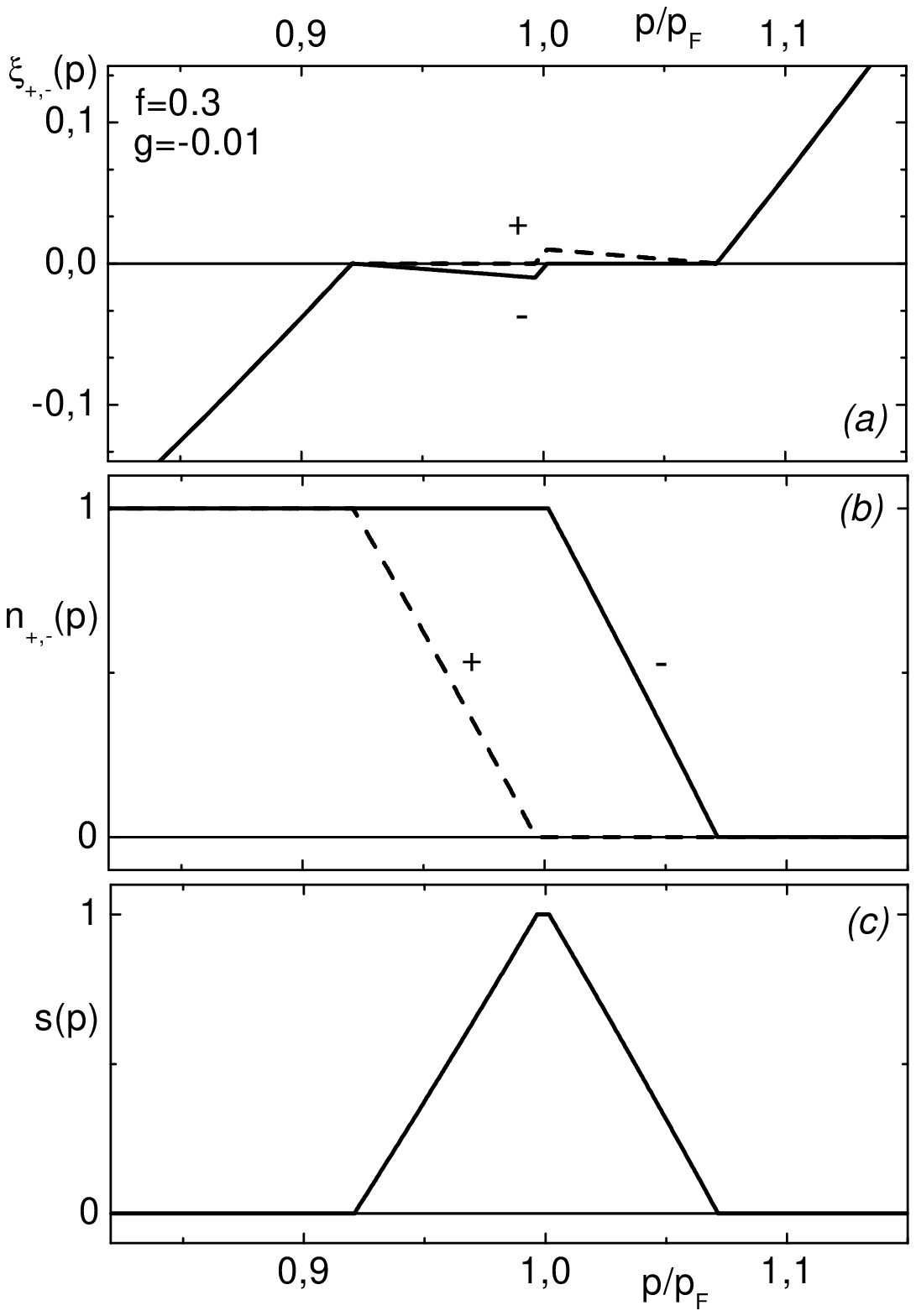,height=8.cm,width=4.5cm}}
\caption{
The same as in Fig.~3 for $f=0.3$, $g=-0.01$.
}
\end{figure}

Some elucidation of the procedure is ensured
by Fig.~2, where the evolution of solutions of the set (\ref{syst2})
versus the input parameter $\xi^0_{{\bf p}}$ is shown. The solid line
corresponds to the dependence
$\xi_+(\xi_-)$, given by the first equation of the set (\ref{syst2}),
while the dashed one shows the same dependence resulting from the
second equation of the set. The intersection points  are indicated by
letters.  Five panels of Fig.~2 show five different cases
referring to different $\xi^0_{\bf p}$. In the first case (panel (a)),
two zigzag lines have one intersection point $A$,
corresponding to the single solution where both
$\xi_+(p)$ and $\xi_-(p)$ are negative.
\begin{figure}
\mbox{\epsfig{file=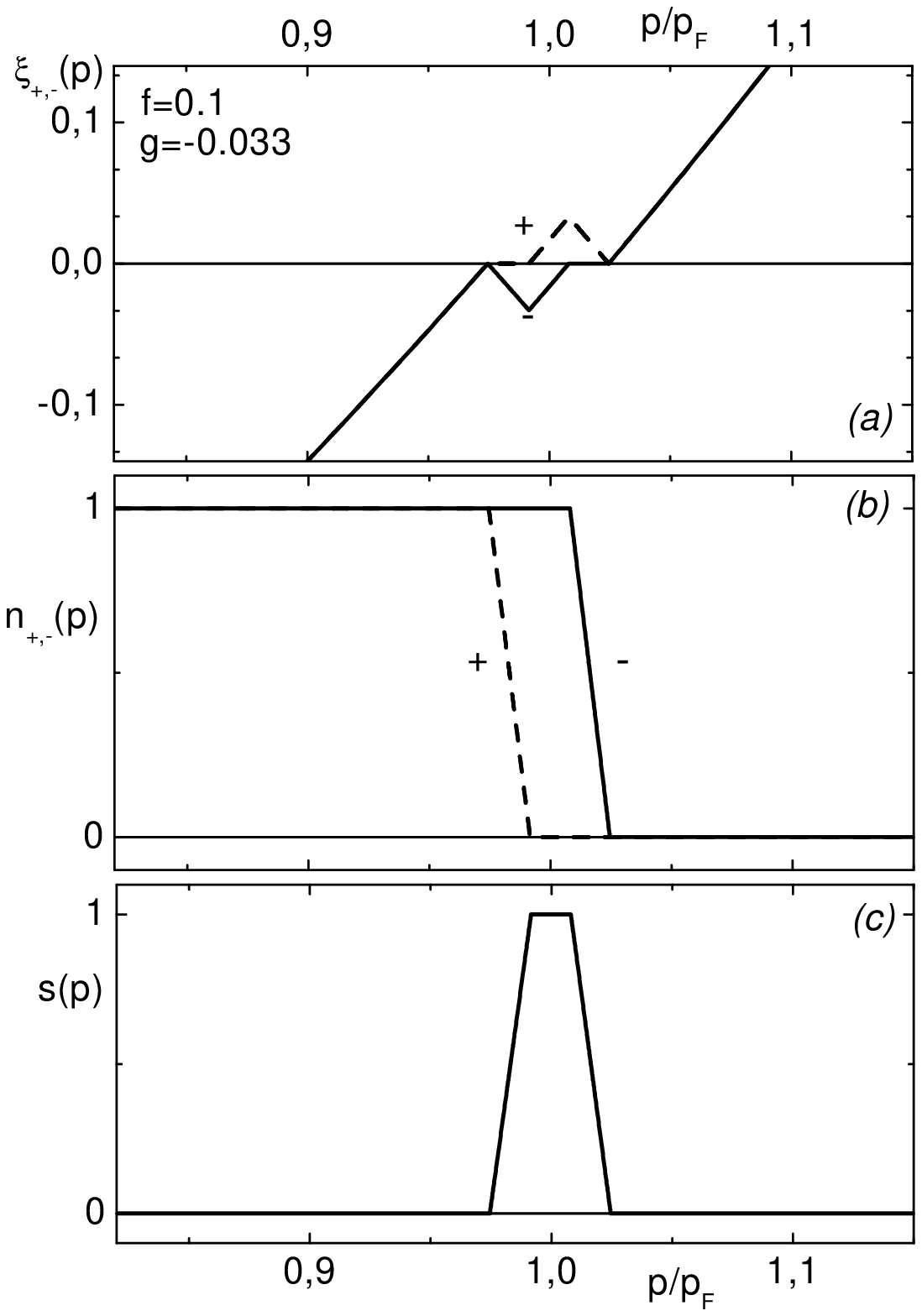,height=8.cm,width=4.5cm}}
\caption{
The same as in Fig.~3 for $f=0.1$, $g=-0.033$.
}
\end{figure}
\begin{figure}
\mbox{\epsfig{file=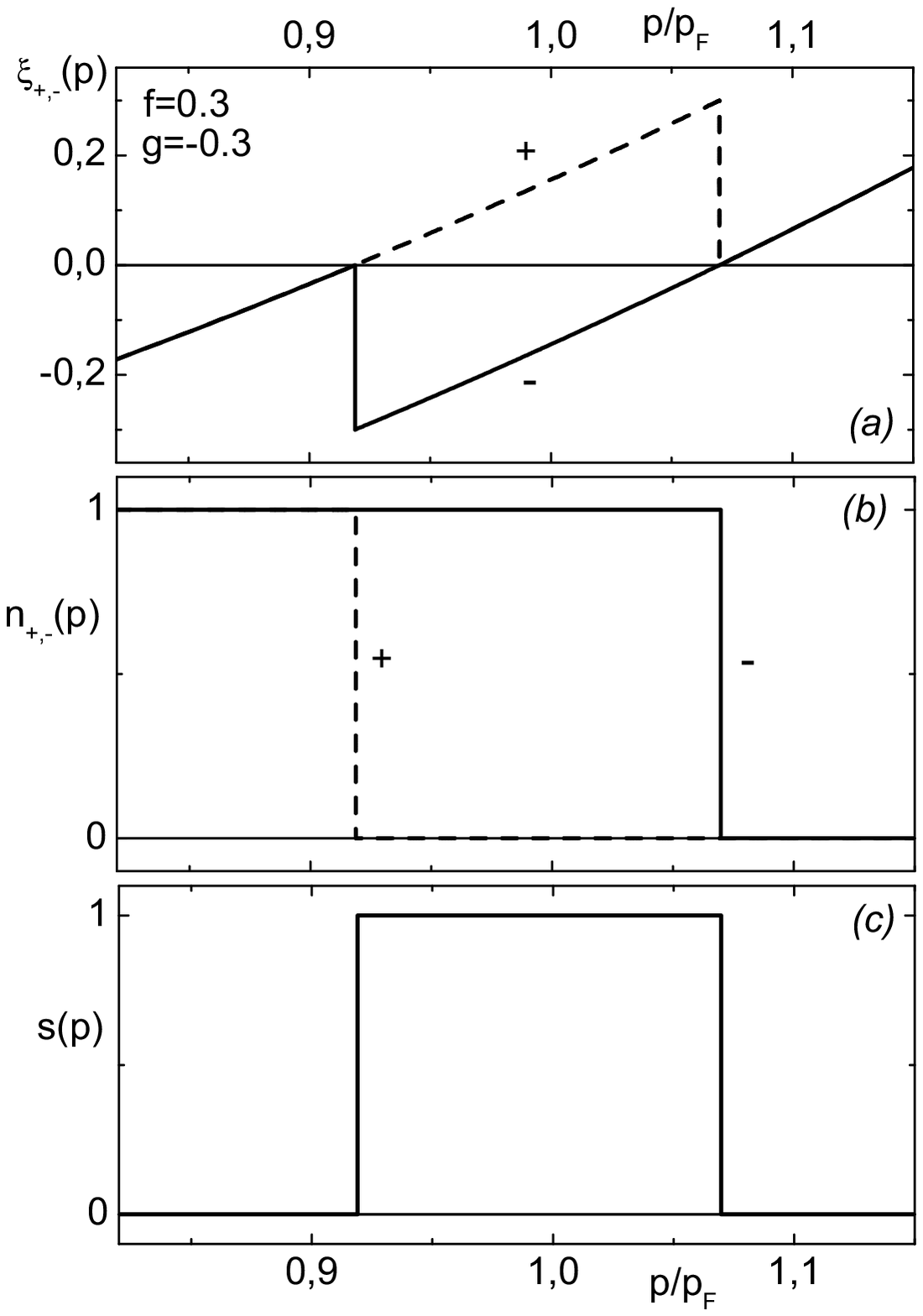,height=8.cm,width=4.5cm}}
\caption{
The same as in Fig.~3 for $f=0.3$, $g=-0.3$.
}
\end{figure}
With increasing $\xi^0_{\bf p}$,  the intersection point
moves to the origin. At a certain value of this variable,
depending on the parameter $f$ only,
bifurcation emerges, and there appear three intersection points
($B_0$, $B_1$ and $B'_1$ on the panel (b)).
One of them, $B_0$, lies at the origin, while the other two,
$B_1$ and $B'_1$, on the $\xi_+$ and $\xi_-$ axes, respectively.
A similar situation occurs in the panels (c) and (d).
\begin{figure}
\mbox{\epsfig{file=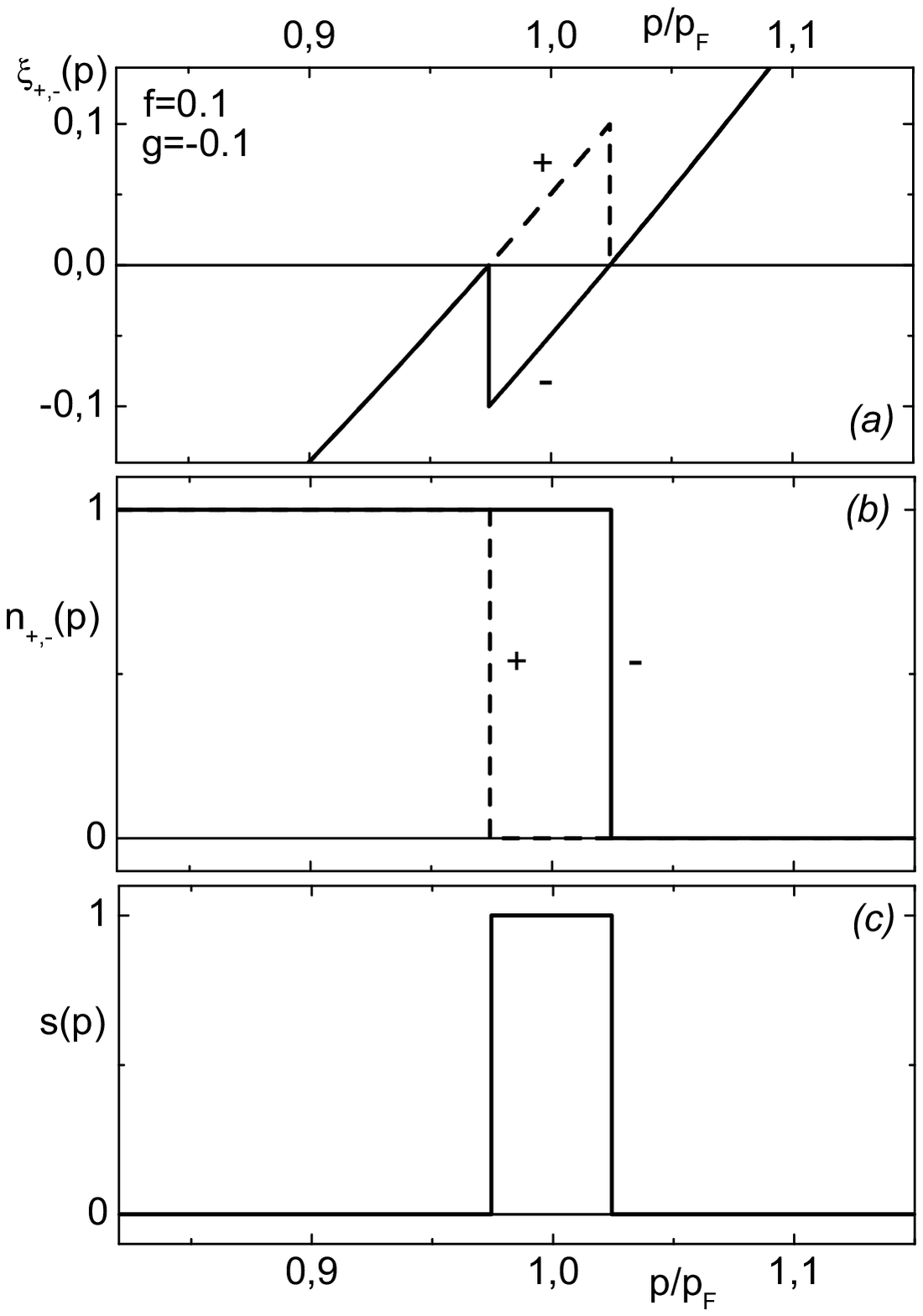,height=8.cm,width=4.5cm}}
\caption{
The same as in Fig.~3 for $f=0.1$, $g=-0.1$.
}
\end{figure}
\begin{figure}
\mbox{\epsfig{file=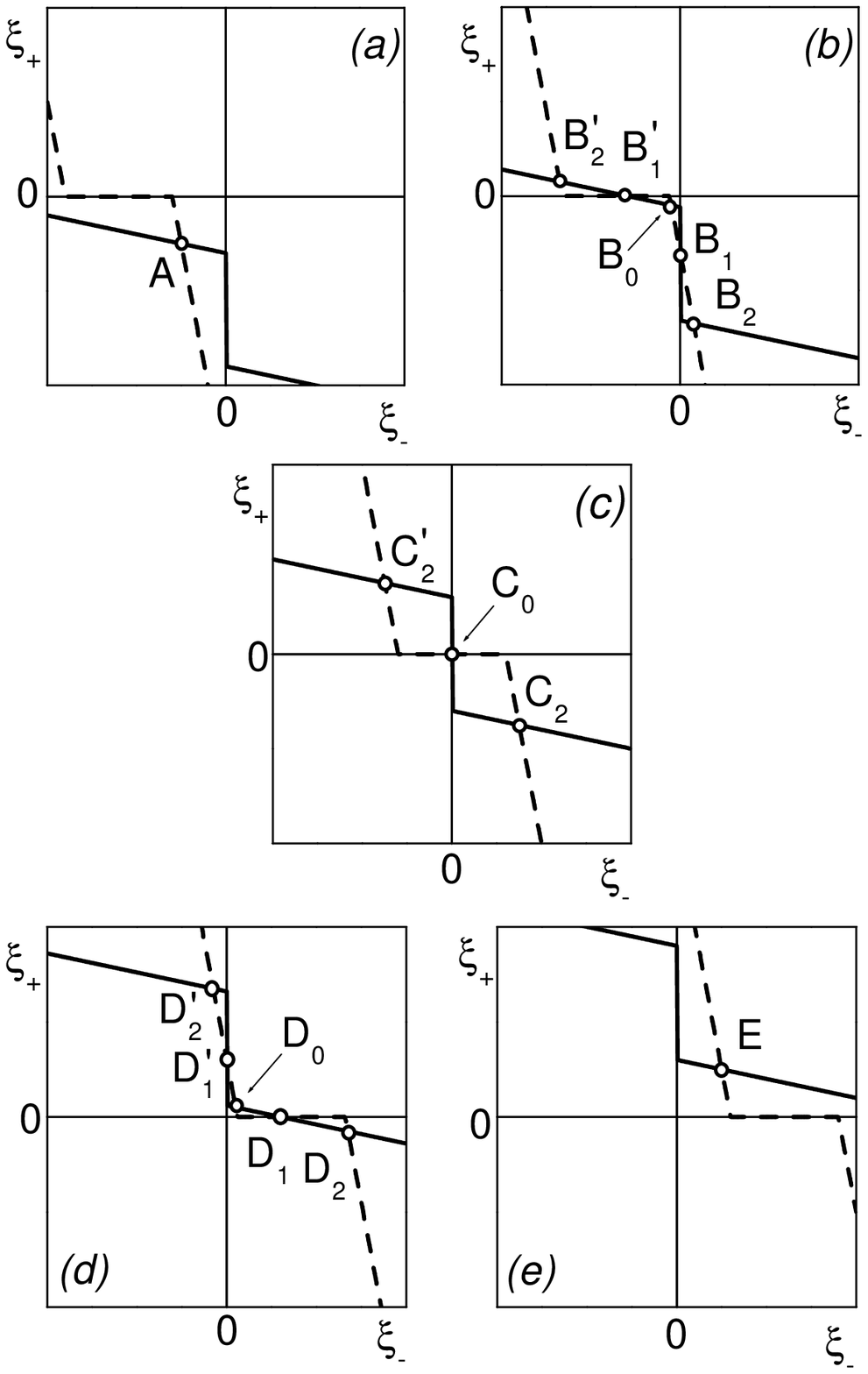,height=9.cm,width=6.cm}}
\caption{
The same as in Fig.~2 for the case of $|g|>f$.
}
\end{figure}
The intersection at the origin persists,
while the other two crossing points lie
inside the quadrants on the panel (c) ($C_1$ and $C'_1$) and
on the $\xi_-$ and $\xi_+$ axes on the panel (d) ($D_1$ and $D'_1$).
The last (e) panel shows the fifth case with one intersection
point at positive values of $\xi_+$ and $\xi_-$.

\begin{figure}
\mbox{\epsfig{file=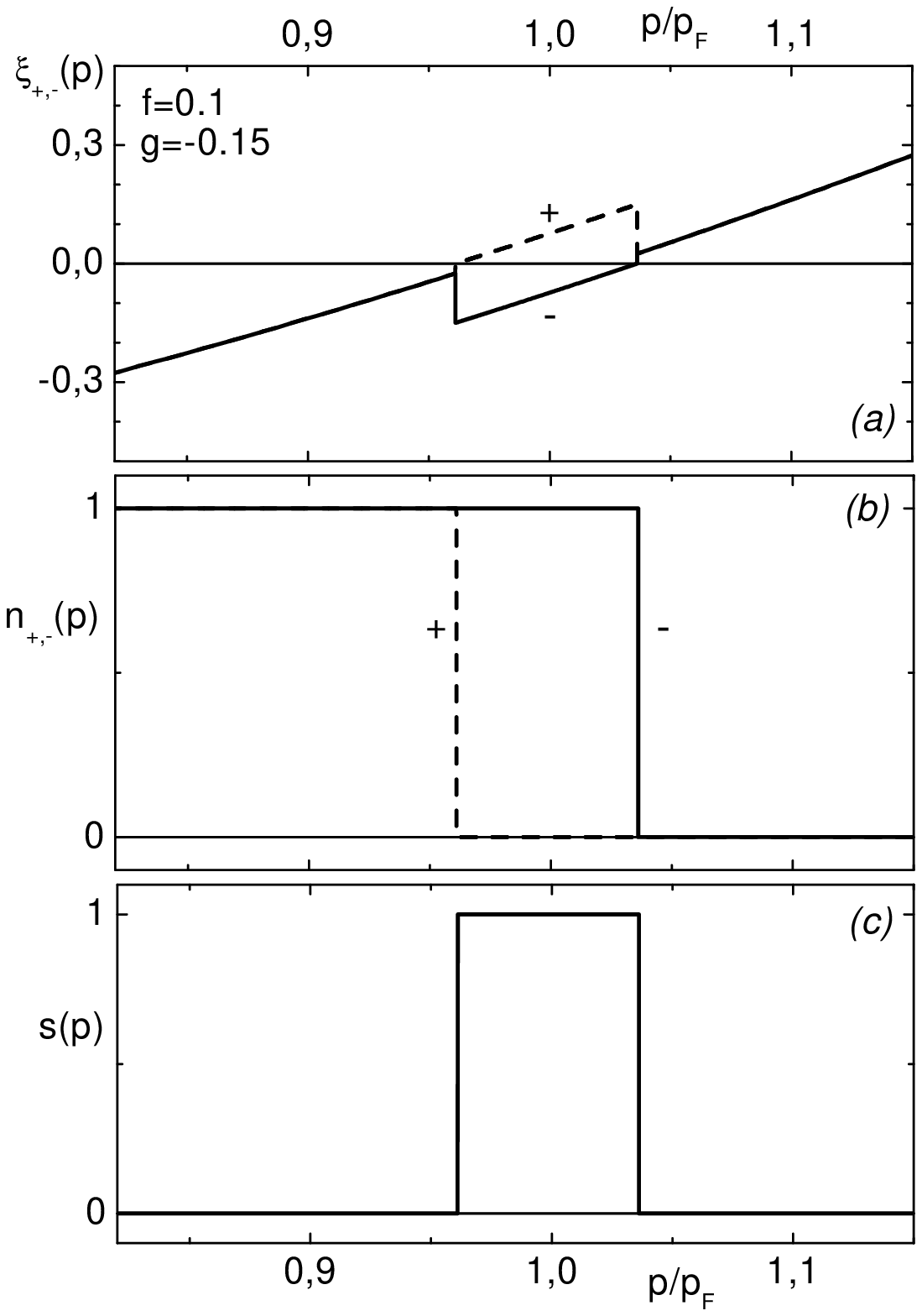,height=8.cm,width=4.5cm}}
\caption{
The same as in Fig.~3 for $f=0.1$, $g=-0.15$.
}
\end{figure}
\begin{figure}
\mbox{\epsfig{file=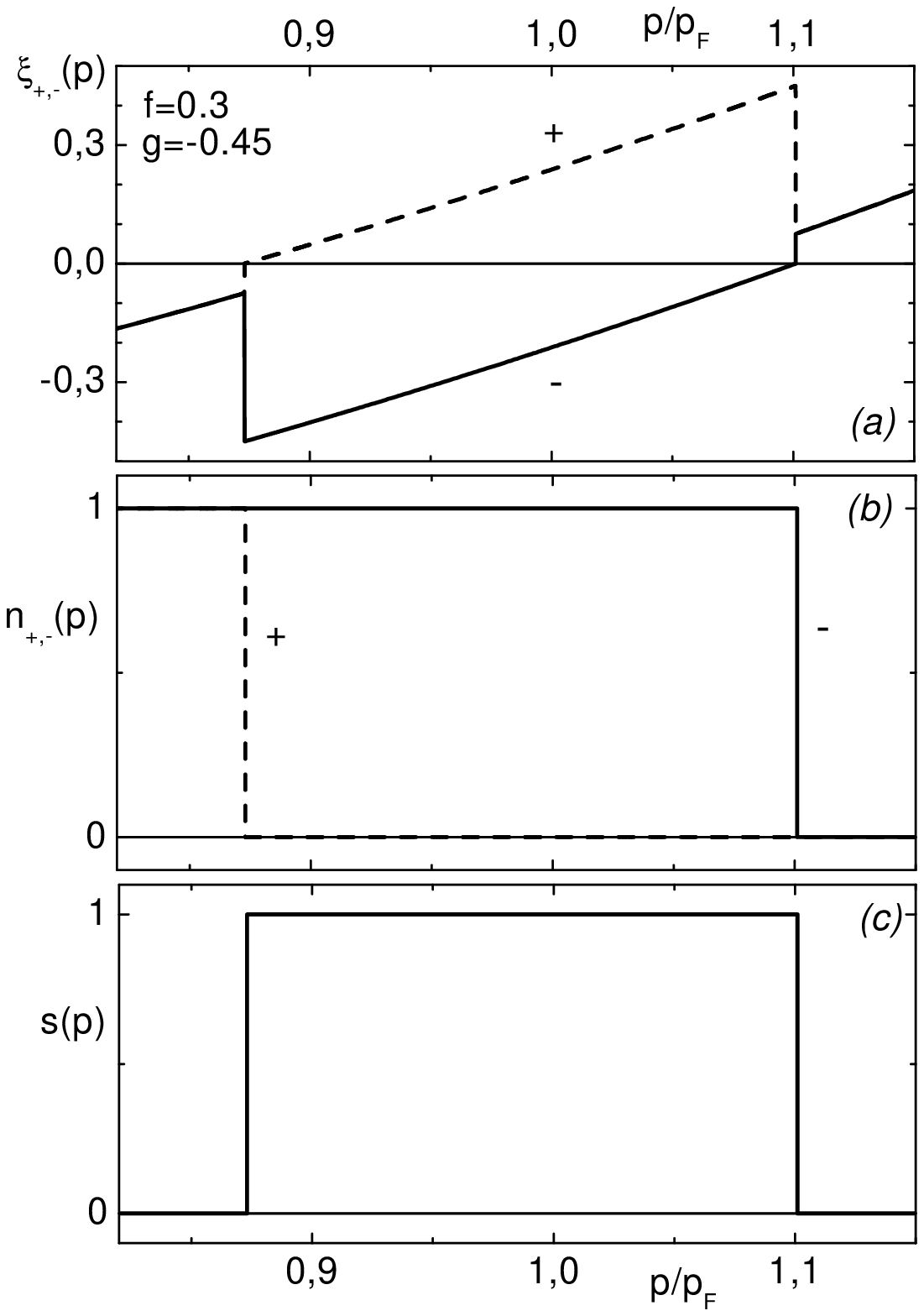,height=8.cm,width=4.5cm}}
\caption{
The same as in Fig.~3 for $f=0.3$, $g=-0.45$.
}
\end{figure}

The chain $A{-}B_0{-}C_0{-}D_0{-}E$ is associated with the solution
(denoted as $\Phi_0$),
for which the sp spectra coincide: $\xi_+(p)=\xi_-(p)=0$  within
the interval $[p_i,p_f]$, the length of which is determined
by the parameter $f$ only, as if there was no spin-spin interaction
at all. The analysis shows that, as long as the constant $g$
remains positive, it is the solution $\Phi_0$ that has the lowest
energy compared with the others.

\begin{figure}
\mbox{\epsfig{file=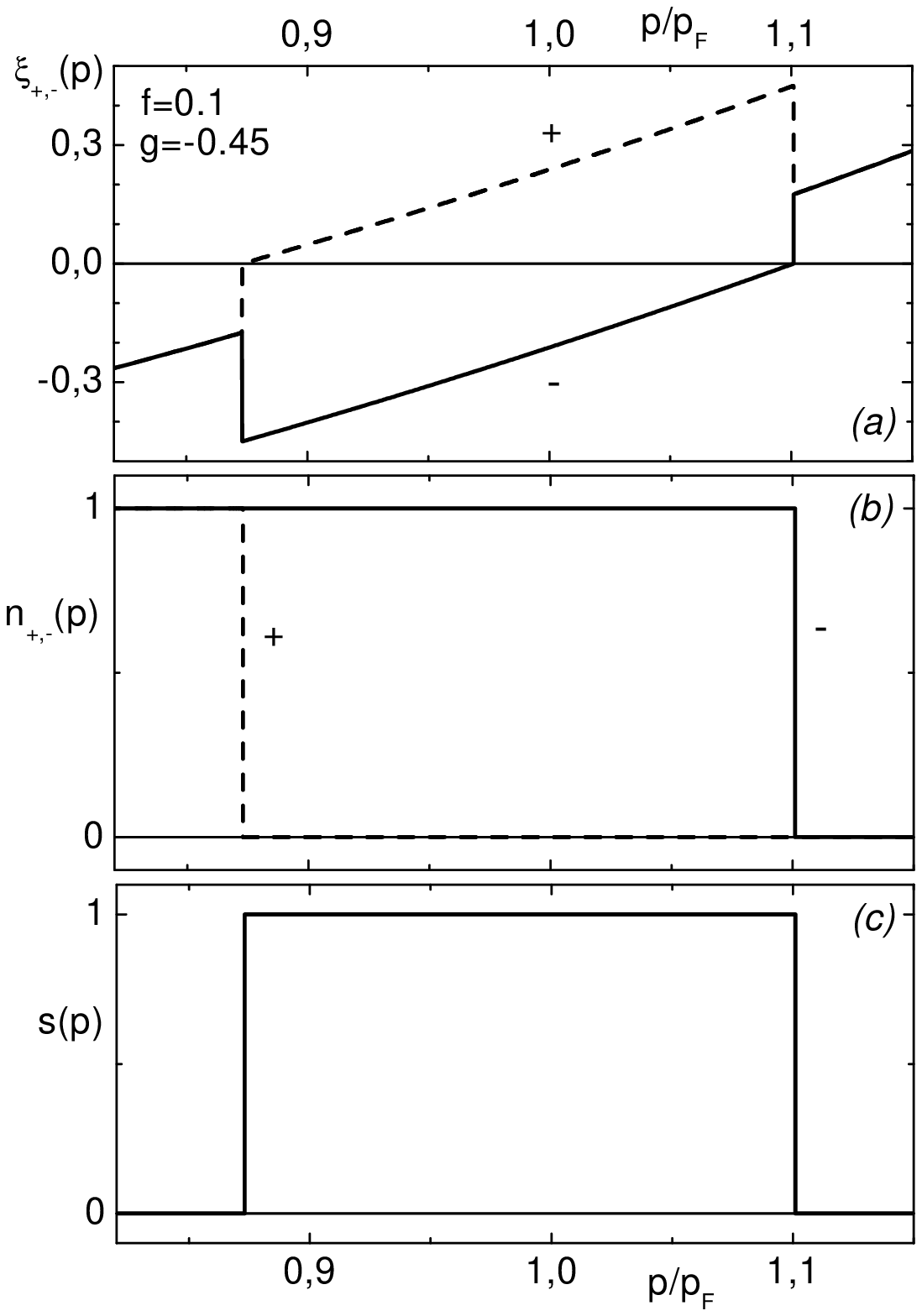,height=8.cm,width=4.5cm}}
\caption{
The same as in Fig.~3 for $f=0.1$, $g=-0.45$.
}
\end{figure}

As for solutions, related to the chains
$A{-}B_1{-}C_1{-}D_1{-}E$ and $A{-}B'_1{-}C'_1{-}D'_1{-}E$
($\Phi_1$ and $\Phi^{\prime}_1$, respectively),
the situation is different.
The solution $\Phi_1$, corresponding to the first chain,
is shown in Fig.~3.
We see that the FC region, the boundaries of which were insensitive
to $g$ at any $g>0$, is destroyed: the spectra $\xi_+(p)$ and $\xi_-(p)$
repell each other. As the momentum $p$ moves from the lower point,
where these spectra simultaneously attain the FS, only the FC plateau
at the spectrum $\xi_+(p)$ survives, with the energy splitting between
the spectra $\xi_+(p)$ and $\xi_-(p)$ growing linearly with
the $p$ increase.
Attaining the maximum at the point, where the plateau at  $\xi_+(p)$
vanishes, this splitting ceases to increase and
remains constant until the point, where a new FC plateau emerges
at the spectrum $\xi_-(p)$, and, finally, both the spectra once
again merge at the point, where the FC disappears forever.

The second chain refers to the solution $\Phi^{\prime}_{1}$,
symmetric to the previous one,
with changing $\pm$ by $\mp$ and, thus, corresponds
to the opposite sign of the projection of spontaneous spin $S$
on the fixed axis. Since both the directions are equivalent,
two solutions with the nonzero $S$ have equal energies. It can be
verified that at any $g<0$, these solutions provide the minimum
of the ground state energy, while the solution $\Phi_0$ with $S=0$,
the maximum. Indeed, calculations yield the gain in energy
of the solutions $\Phi_1$ and $\Phi^{\prime}_{1}$,
as compared with the energy of the Landau state
$E_1/E_L-1\simeq -0.027$, while for the solution $\Phi_0$,
one obtains $E_0/E_L-1\simeq -0.019$.

Thus, the reason for these alterations of the FC
is the occurrence of the spontaneous spin
\beq
S=\sum\limits_p s(p)=\sum\limits_p \biggl(n_-(p)-n_+(p)\biggr)  \  .
\label{spon}
\eeq
arising at any $g<0$. For illustration, the spin density $s(p)$
is drawn in panel (c).
We infer that the spin density $s(p)$ differs from 0 only
in the region $[p_i,p_f]$. Therefore the total spin $S$
depends primarily
on the value of parameter $f$. For the parameters
$f$ and $g$, given in Fig.~3, one obtains $S/\rho\simeq 0.299$.

For comparison, results for the other two sets of the parameters:
$f=0.3$, $g=-0.01$ and $f=0.1$, $g=-0.033$ are drawn in Figs.~4
and 5, respectively. The striking feature of these results is that
for the given value $f=0.3$ the magnitude of the spontaneous spin
$S$ for $g=-0.01$, $S/\rho\simeq 0.232$, is of the same order as
in the case of $g=-0.1$. The energy gain of the solution with the
spontaneous spin, corresponding to $f$ and $g$ for Fig.~4,
$E_1/E_{L}-1\simeq -0.019$. Fig.~5 shows that multiplication of
both the parameters $f$ and $g$ by the factor $1/3$ results
in the scaling of the spectra, occupation numbers, and spontaneous
spin by the same factor.

The situation drastically changes if the magnitude of the
spin-spin force becomes comparable to that of the scalar one. This
is seen in Figs.~6 and 7 showing results obtained for the
sets $f=0.3$, $g=-0.3$ and $f=0.1$, $g=-0.1$. We see that the FC
practically disappears, while the Landau state reappears. Now the
occupation numbers $n_+(p)$ and $n_-(p)$ have the Migdal jump from
0 to 1 at the points $p_i$ and $p_f$, respectively, and the
spontaneous spin attains the maximum value, which is equal to the
phase volume of the spherical layer between $p_i$ and $p_f$
($S/\rho\simeq 0.448$ and $S/\rho\simeq 0.149$, for the parameters
of Figs.~6 and 7, respectively). Fig.~8 illustrates the evolution
of solutions for the set (\ref{syst2}) versus $\xi^0_{{\bf p}}$ in
the case when the absolute value of the spin-spin constant $g$ is
larger than the scalar one $f$. In this case, the set
(\ref{syst2}) has five solutions: $\Phi_0$, corresponding to
the chain $A{-}B_0{-}C_0{-}D_0{-}E$; $\Phi_1$ and
$\Phi^{\prime}_1$, corresponding to the chains
$A{-}B_1{-}C_0{-}D_1{-}E$ and $A{-}B'_1{-}C_0{-}D'_1{-}E$,
respectively; and a new couple of solutions, $\Phi_2$ and
$\Phi^{\prime}_2$, corresponding to the chains
$A{-}B_2{-}C_2{-}D_2{-}E$ and $A{-}B'_2{-}C'_2{-}D'_2{-}E$. The
calculation of the energies yields the solutions $\Phi_2$ and
$\Phi^{\prime}_2$, equal in energy, beating $\Phi_0$, $\Phi_1$
and $\Phi^{\prime}_1$. The solution $\Phi_2$ is
drawn in Figs.~9, 10, 11 for three sets of parameters: $f=0.1$,
$g=-0.15$; $f=0.3$, $g=-0.45$; $f=0.1$, $g=-0.45$. These figures
show that in this case the spontaneous spin is defined only by
the parameter $g$, and the parameter $f$ defines the magnitudes of
the jumps in the spectra $\xi_+(p)$ and $\xi_-(p)$. It is worth
noting that, while the flat portions of the spectra, corresponding
to FC, disappear and the Landau state is recovered, the phantom of
the FC manifests itself in the emergence of the spontaneous spin.

\begin{figure}
\mbox{\epsfig{file=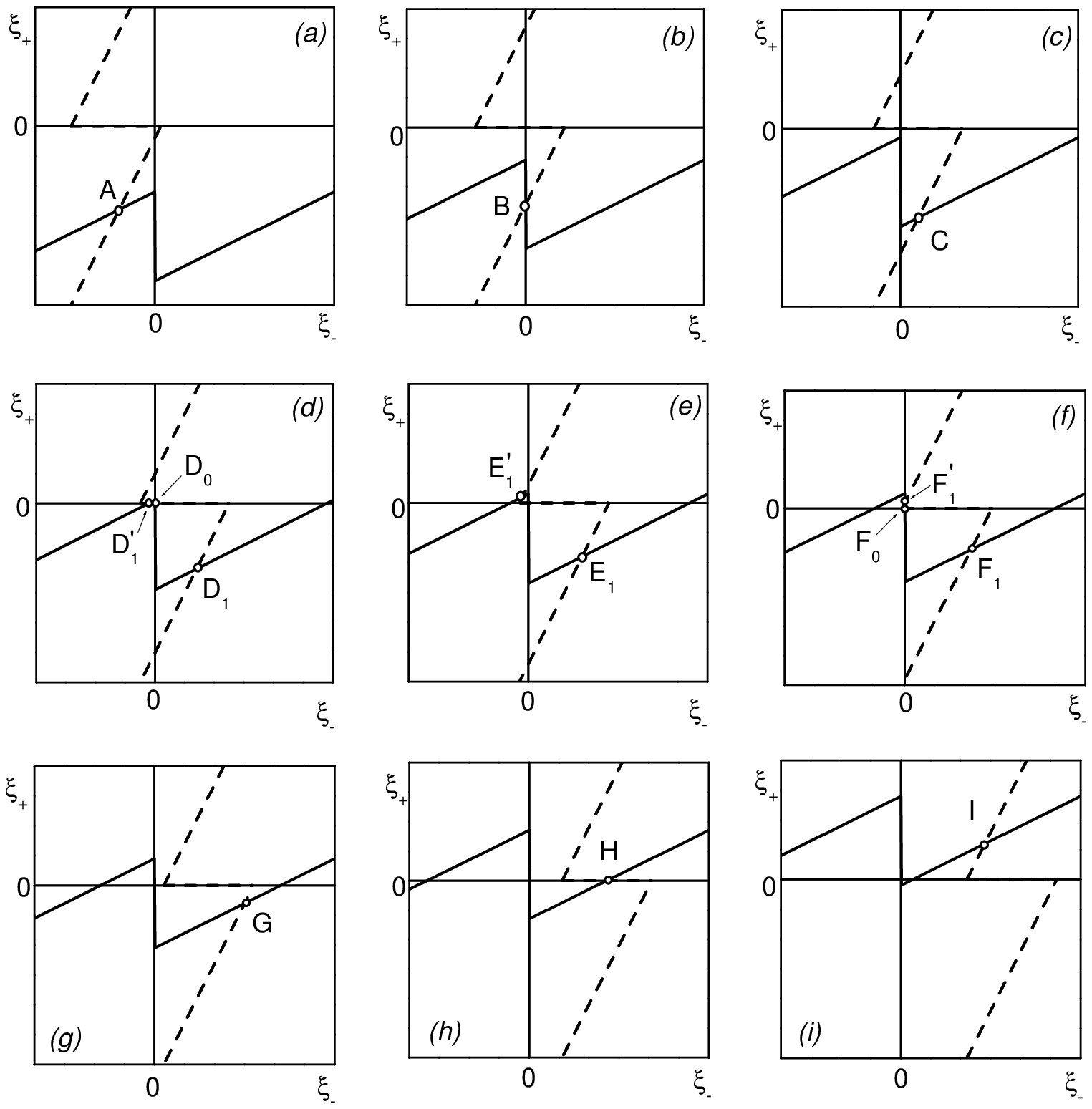,height=8.cm,width=8.cm}}
\caption{
Graphical illustration of the dynamics of solutions of
the set (\ref{systh2}) with increasing $\xi^0_{\bf p}$
in the case of $|g|<f$, $\beta H < |g|/2$.
}
\end{figure}

The above analysis can be generalized to the case of an external
magnetic field. In this case, the equations for the sp
spectra read
\begin{eqnarray}
\xi_+(p)&=&\xi^0_{{\bf p}} +{1\over 2}f(n_+(p)+n_-(p))
\nonumber \\ &&
+{1\over 2}g(n_+(p)-n_-(p)) + \beta H \ ,
\nonumber \\
\xi_-(p)&=&\xi^0_{{\bf p}} +{1\over 2}f(n_+(p)+n_-(p))
\nonumber \\ &&
-{1\over 2}g(n_+(p)-n_-(p)) - \beta H \ ,
\label{magn}
\end{eqnarray}
where $H$ is the effective magnetic field acting on the
fermion spin and $\beta$ is the magnetic moment of the fermion.
Upon rewriting Eqs.~(\ref{magn}) to the form
convenient for the graphical analysis, we obtain
\begin{eqnarray}
\xi_+=\left(1{-}{a\over b}\right)\xi^0_{{\bf p}}
+{a\over b}\xi_- + \left(b{-}{a^2\over b}\right)n_-
+ \left(1{+}{a\over b}\right)\beta H ,
\nonumber \\
\xi_-=\left(1{-}{a\over b}\right)\xi^0_{{\bf p}}
+{a\over b}\xi_+ + \left(b{-}{a^2\over b}\right)n_+
- \left(1{+}{a\over b}\right)\beta H .
\label{systh2}
\end{eqnarray}

\begin{figure}
\mbox{\epsfig{file=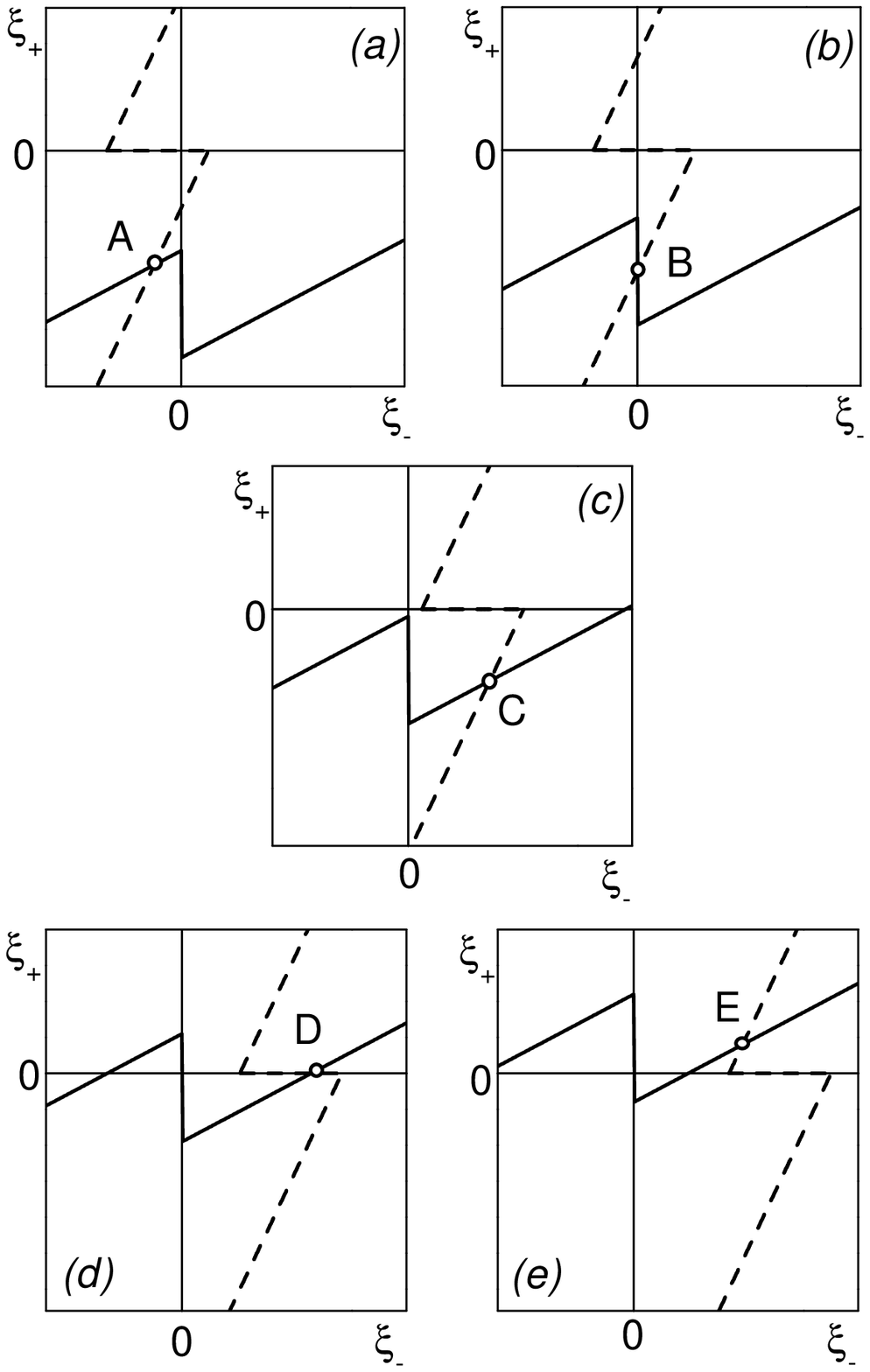,height=9.cm,width=6.cm}}
\caption{
The same as in Fig.~12 for the case of
$\beta H > |g|/2$.
}
\end{figure}
\begin{figure}
\mbox{\epsfig{file=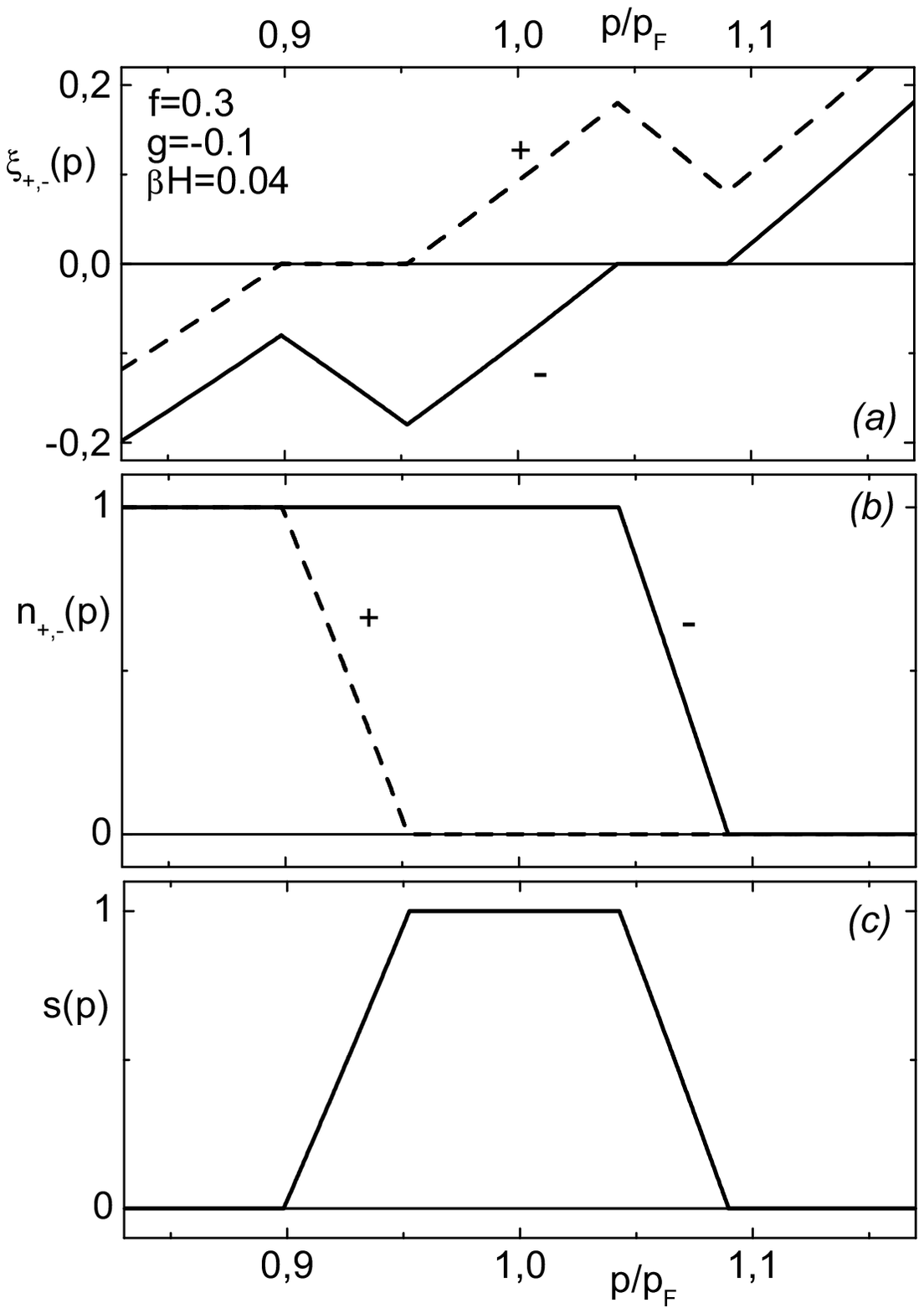,height=8.cm,width=4.5cm}}
\caption{
The same as in Fig.~3 for $f=0.3$, $g=-0.1$,
$\beta H=0.04$.
}
\end{figure}

Solutions of Eqs.~(\ref{systh2}) are represented by the broken
lines with the
same slopes and jumps as in the case of zero magnetic field
but shifted by the value $(1+a/b)\beta H$ along the $\xi_+$
and $\xi_-$ axes. For $|g|<f$, two cases can be distinguished.
In the case of $\beta H<|g|/2$ ($\beta H$ is in units of
$\varepsilon^0_F$), drawn in Fig.~12, two segments, vertical and
horizontal, lying on the axes, intersect each other and
the situation with the three intersection points, discussed
above, holds.
\begin{figure}
\mbox{\epsfig{file=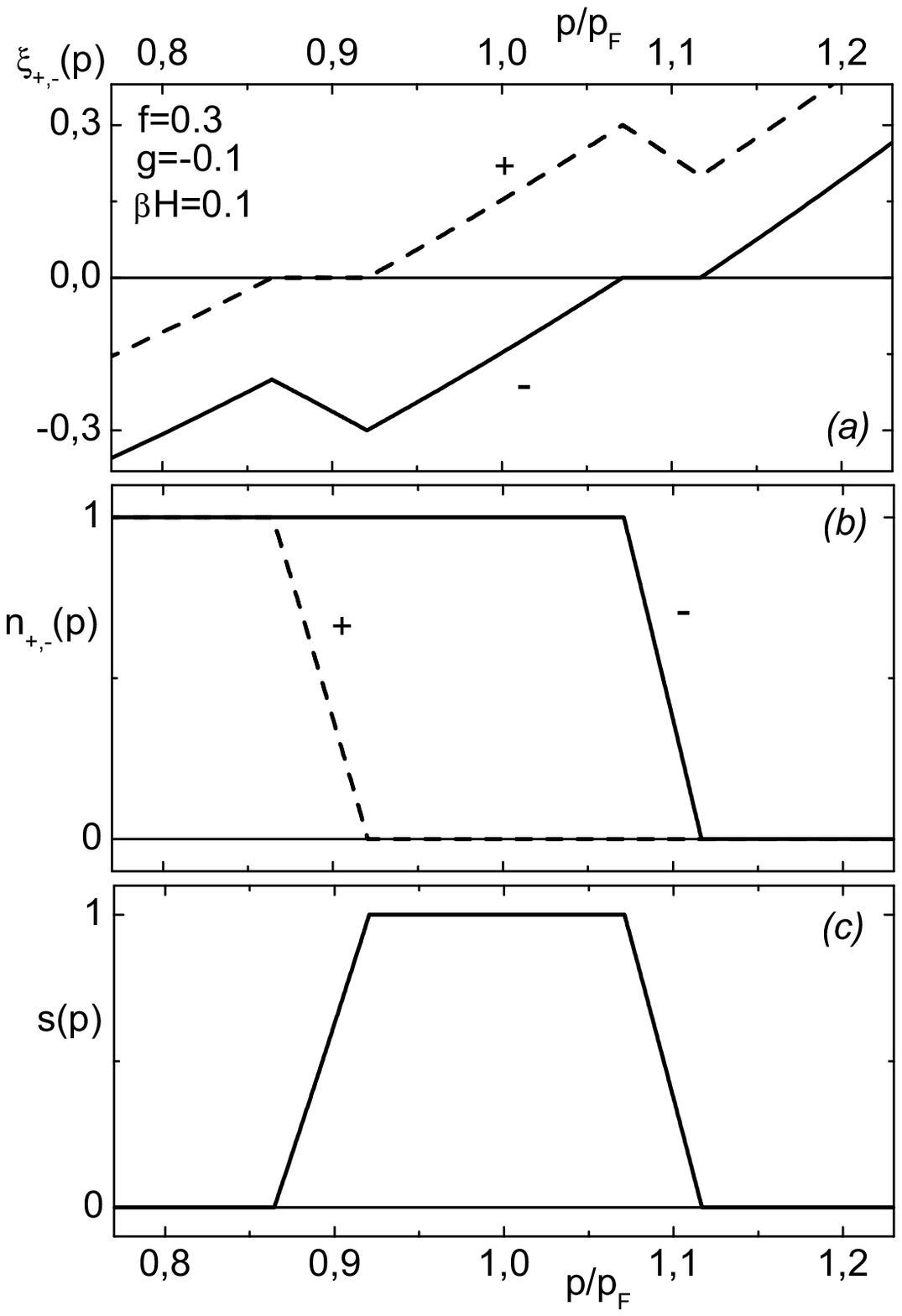,height=8.cm,width=4.5cm}}
\caption{
The same as in Fig.~3 for $f=0.3$, $g=-0.1$,
$\beta H=0.1$.
}
\end{figure}
In the case $\beta H>|g|/2$, the single intersection
point of the two lines remains at any $\xi_0$ (see Fig.~13).
The sp spectra and occupation numbers are shown
in Fig.~14 for $f=0.3$, $g=-0.1$, $\beta H=0.04$
and in Fig.~15 for the same values of $f$ and $g$
but at $\beta H=0.1$. In both the cases, the magnetic
field promotes splitting the sp spectra $\xi_+(p)$ and $\xi_-(p)$
but does not influence their flat parts.

In conclusion, in the Nozieres-like model of a Fermi system
with scalar and spin-dependent long-range forces, it is
shown that the fermion condensation, occurring in the vicinity
of a phase transition, results in the emergence of the
weak magnetization which precedes the ferromagnetic transition.

\begin{acknowledgments}
We thank A.~Lichtenstein, G.~Kotliar and V.~Yakovenko
for many valuable discussions.

This research was supported in part by the National Science
Foundation under Grants No. PHY99-07949 and No. PHY-0140316,
 by the McDonnell Center
for the Space Sciences at Washington University,
and by Grant No.~NS-1885.2003.2 from the Russian Ministry
of Industry and Science (VAK and MVZ).  One of the authors (VAK) thanks
the University of California
(Santa Barbara) for the kind hospitality.
\end{acknowledgments}

\end{document}